\documentclass[aps,prl,floatfix,showpacs,twocolumn,superscriptaddress]{revtex4}
\usepackage{amsmath,amssymb}
\usepackage{graphicx}
\usepackage{epsfig}
\usepackage{psfrag}
\usepackage[usenames]{color}

\begin{document}

\hsize\textwidth\columnwidth\hsize\csname@twocolumnfalse\endcsname

\title{Paramagnetic Resonance in Spin-Polarized Disordered Bose Condensates}

\author{V. M. Kovalev}
\email{vadimkovalev@isp.nsc.ru}
\affiliation{Center for Theoretical Physics of Complex Systems, Institute for Basic Science, Daejeon, Republic of Korea}
\affiliation{Institute of Semiconductor Physics, Siberian Branch of Russian Academy of Sciences, Novosibirsk 630090, Russia}
\affiliation{Department of Applied and Theoretical Physics, Novosibirsk State Technical University, Novosibirsk 630073, Russia}
\author{I. G. Savenko}
\affiliation{Center for Theoretical Physics of Complex Systems, Institute for Basic Science, Daejeon, Republic of Korea}
\affiliation{National Research University of Information Technologies, Mechanics and Optics, St. Petersburg 197101, Russia}
\affiliation{Nonlinear Physics Centre, Research School of Physics and Engineering, The Australian National University, Canberra ACT 2601, Australia}


\date{\today}

\begin{abstract}
We study the pseudo-spin density response of a disordered two-dimensional spin-polarized Bose gas of exciton polaritons to weak alternating magnetic field, assuming that one of the spin states of the doublet is macroscopically occupied and Bose-condensed while the occupation of the other state remains much smaller. We calculate spatial and temporal dispersions of spin susceptibility of the gas taking into account spin-flip processes due to the transverse-longitudinal splitting. Further, we use the Bogoliubov theory of weakly-interacting gases and show that the time-dependent magnetic field power absorption exhibits double resonance structure corresponding to two particle spin states (contrast to paramagnetic resonance in regular spin-polarized electron gas). We analyze the widths of these resonances caused by scattering on the disorder and show that, in contrast with the ballistic regime, in the presence of impurities, the polariton scattering on them is twofold:  scattering on the impurity potential directly and scattering on the spatially fluctuating condensate density caused by the disorder. As a result, the width of the resonance associated with the Bose-condensed spin state can be surprizingly narrow in comparison with the width of the resonance associated with the non-condensed state.


\pacs{76.30.-v,71.35.Gg,71.36.+c}


\end{abstract}

\maketitle
\textit{Introduction.---}
Due to their hybrid half-light--half-matter nature, microcavity exciton polaritons (EPs) demonstrate a number
of peculiar properties, standing aside from other quasiparticles in solid-state. In particular, their small effective mass ($10^{-4}-10^{-5}$ of free electron mass) and relatively short lifetime ($5$-$100$ ps) inherited from the photons together with strong particle-particle interaction taken from the excitons make EP systems suitable for observation of quantum collective phenomena at astonishingly high temperatures~\cite{KasprzakNature, Christopoulos}. Other significant effects have been reported, such as EP superfluidity~\cite{AmoNature}, the Josephson effect~\cite{LagoudakisJosephson}, formation of vortices~\cite{Vortex}. Some of the theoretically predicted phenomena such as polariton self-trapping~\cite{ShelykhJosephson}, polariton-mediated superconductivity~\cite{LaussySupercond} are to be tested and confirmed. Indeed, from the fundamental viewpoint, EPs represent a testbed which role cannot be overestimated.

Beside fundamental importance, semiconductor microcavities operating in the strong coupling regime can be used in various optoelectronic applications~\cite{ReviewPolaritonDevices}. Certainly, a polariton laser should be mentioned here~\cite{Imamoglu, RefNature, RefKarpovAPL} as a manifestation of BEC-based alternative light source. Coherently pumped microcavities also give us polariton neurons~\cite{LiewNeuron} and
polariton integrated circuits~\cite{LiewCircuit}. Further, semiconductor microcavities under incoherent background pumping (for instance, electric current injection) can be used in optical routers~\cite{OurAPL, BlochAPL}, detectors of terahertz radiation~\cite{RefTHzAPL, RefTHzPRL}, high-speed optical switches~\cite{Nat4,Wertz2010} and more.

One of the most intriguing and significant quantum properties governing the dynamics of EPs, is their spin degree of freedom (also referred to as polarization)~\cite{RefShelykhSpin}. It opens a way to spin-optronics~\cite{ShelykhSpinoptronics}. One one hand, as opposed to classical optics, where nonlinear Kerr interaction is usually weak, spin-optronics is in a more favourable position thank to advantageous relatively strong particle-particle interaction. On the other hand, as opposed to spintronics, using EPs can reduce the dramatic impact of the carrier spin relaxation and decoherence~\cite{RefGlazov, RefGlazov2, Glazov, SpinHall}.

Polariton spin dynamics has been extensively studied in literature~\cite{Dufferwiel, Ohadi, FlayacHalfSol, Tercas2013}, although many issues remain undiscovered. For example, from the physics of electrons in metals we know such effect as the \textit{paramagnetic resonance} also called the electron spin resonance~\cite{Zavoisky}.
In our manuscript we study the paramagnetic resonance in a semiconductor microcavity under time-dependent magnetic field. We show that in the case of EP Bose condensate in the presence of a disorder, paramagnetic resonance has several peculiarities, unusial for two-dimensional (2D) electronic systems.

\textit{Pseudospin susceptibility.---}
Dynamics of EPs in a microcavity can be described by the spinor wave function, having two components related to two polariton spin states,
$\hat{\psi}(\textbf{r},t)=(\psi^\dagger_+(\textbf{r},t),\psi^\dagger_-(\textbf{r},t))^T$.
Our goal is to study the response of the polariton spin density $S^\alpha(\textbf{r},t)=\hat{\psi}^\dagger(\textbf{r},t)\sigma^\alpha\hat{\psi}(\textbf{r},t)$
to external space and time fluctuating magnetic field in the Faraday configuration, $\textbf{B}(\textbf{r},t)=(0,0,B(\textbf{r},t))$. Let us use a usual restriction: assume that the magnitude of this field is low enough thus a linear response theory can be applied.
In its framework, the spin susceptibility is defined as
\begin{gather}\label{eq1}
S^i(\textbf{r},t)=\iint d\textbf{r}' dt'\chi_{ij}(\textbf{r},\textbf{r}';t,t')B_j(\textbf{r}',t').
\end{gather}
First, we should describe the EP dynamics by the Gross-Pitaevskii equation (GPE), accounting for the polariton-impurity, polariton-polariton interaction and the TE-TM spin splitting. Utilizing the interacting Hamiltonian in a special form~\cite{RuboBookChapter},
$$\hat H_\textrm{int}=\frac{1}{2}U_0\left(|\psi_+|^4+|\psi_-|^4\right)+U_2|\psi_+|^2|\psi_-|^2,$$
where $U_2=U_0-2U_1$, $U_0$ and $U_1$ are polariton-polariton interaction constants, we can write the GPE for each of the spin components of the EP doublet:
%
\begin{gather}\label{eq2}
i\dot{\psi}_\pm=\left(\hat{E}_{\textbf{p}}-\mu+u(\textbf{r})+U_0|\psi_\pm|^2~~~~~~~~~~~~~~~\right.\\
\nonumber
\left.~~~~~~~~~~~~~+U_2|\psi_\mp|^2\pm\mathcal{F}\right)\psi_\pm+\alpha p_\mp^2\psi_\mp,
\end{gather}
%
where $\hat{E}_{\textbf{p}}=\hat{\textbf{p}}^2/2M$ is the operator of kinetic energy of EPs with the mass $M$ (we assume parabolic dispersion at not very high $p$ for simplicity). The non-diagonal terms $\alpha p^2_{\pm}=\alpha(p_x\pm ip_y)^2$
account for the TE-TM splitting of polariton states, mixing the $'+'$ and $'-'$ spinor
components. An external magnetic perturbation is given here via the term
$\mathcal{F}(\textbf{r},t)=\frac{1}{2}g_s\mu_bB(\textbf{r},t)$.
Here $g_s$ is an effective polariton $g$-factor, $\mu_b$ is the Bohr magneton, and we also assume that
the perturbation is real for simplicity, $B^*(\textbf{r},t)=B(\textbf{r},t)$. Randomly fluctuating impurity potential
is assumed to have zero mean value, $\langle u(\textbf{r})\rangle=0$, and the following statistical properties:
\begin{gather}\label{eq3}
\langle u(\textbf{r})u(\textbf{r}')\rangle=u_0^2\delta_{\textbf{r},\textbf{r}'},\,
\langle u(\textbf{p})u(\textbf{p}')\rangle=u_0^2\delta_{\textbf{p},\textbf{p}'},
\end{gather}
where $\langle...\rangle$ means the averaging over the impurities positions.

In the steady state (quasi-equilibrium) and in the absence of TE-TM splitting, the ground state of the EP condensate is sensitive to the sign of the interacting parameter, $U_1$~\cite{RuboBookChapter, RefShelykhSpin}. If $U_1>0$, the ground state is a composition of equally populated spin-up and spin-down components of EP spinor. If, instead, $U_1<0$, the ground state is characterized by nearly zero population of one of the circular component of the EP spinor and macroscopic population of the other one, we consider the latter case.
Under the action of external perturbation, $\mathcal{F}(\textbf{r},t)$, the TE-TM terms cause transitions of EPs from the condensed component (let it be $\psi_+$) to the other one ($\psi_-$), which was empty initially. We assume that the occupation of the condensed component ever remains much larger, $|\psi_+|^2\gg|\psi_-|^2$. Then we can disregard the non-linear terms proportional to $U_0|\psi_-|^2$ and $(U_0-2U_1)|\psi_-|^2$ in Eq.~(\ref{eq2}).
%
%
After these agreements, we can write the system of equations which describes evolution of the spinor components of the EP field:
\begin{gather}\label{eq4}
\left(i\partial_t-\hat{E}_{\textbf{p}}+\mu-U_0|\psi_+|^2-u(\textbf{r})-\mathcal{F}\right)\psi_+=\alpha p_-^2\psi_-,\\
\nonumber
\left(i\partial_t-\hat{E}_{\textbf{p}}+\mu-U_2|\psi_+|^2-u(\textbf{r})+\mathcal{F}\right)\psi_-=\alpha p_+^2\psi_+.
\end{gather}
Considering here $\mathcal{F}$ as a perturbation, we find the solution in the form of non-perturbed terms and small corrections,
\begin{gather}\label{eq7}
\left(
  \begin{array}{c}
    \psi_+(\textbf{r},t) \\
    \psi_-(\textbf{r},t) \\
  \end{array}
\right)
\rightarrow\left(
  \begin{array}{c}
    \psi_0(\textbf{r})+\delta\psi_+(\textbf{r},t) \\
    \delta\psi_-(\textbf{r},t) \\
  \end{array}
\right),
\end{gather}
where we have extracted the condensate fraction, $\psi_0(\textbf{r})$, of $\psi_+$ polariton state and introduced new symbol for the non-condensed component, $\psi_-\rightarrow\delta\psi_-$, assuming $\delta\psi_+\sim\delta\psi_-\sim\mathcal{F}$.
Substituting~(\ref{eq7}) into~(\ref{eq4}) and keeping only zero and first-order terms with respect to $\mathcal{F}$, we find that zero-order terms describe the ground state of EP condensate in the impurity potential:
\begin{gather}\label{eq8}
[\hat{E}_{\textbf{p}}-\mu+U_0|\psi_0(\textbf{r})|^2+u(\textbf{r})]\psi_0(\textbf{r})=0,
\end{gather}
while the first-order terms contain information about EP dynamics due to external perturbations,
\begin{gather}\label{eq9}
\hat{G}^{-1}\left(
         \begin{array}{c}
           \delta\psi_+ \\
           \delta\psi_+^* \\
         \end{array}
       \right)-\hat{K}\left(
         \begin{array}{c}
           \delta\psi_- \\
           \delta\psi_-^* \\
         \end{array}
       \right)=\psi_0(\textbf{r})\mathcal{F}(\textbf{r},t)\left(
                                          \begin{array}{c}
                                            1 \\
                                            1 \\
                                          \end{array}
                                        \right),\\\nonumber
\hat{\mathfrak{G}}^{-1}\left(
         \begin{array}{c}
           \delta\psi_- \\
           \delta\psi_-^* \\
         \end{array}
       \right)-\hat{K}^*\left(
         \begin{array}{c}
           \delta\psi_+ \\
           \delta\psi_+^* \\
         \end{array}
       \right)=0,\,\,\,\hat{K}=\left(
                 \begin{array}{cc}
                   \alpha p_-^2 & 0 \\
                   0 & \alpha p_+^2 \\
                 \end{array}
               \right),
\end{gather}
and the Green's functions introduced in Eq.~(\ref{eq9}) read:
\begin{widetext}
\begin{gather}
\nonumber
\hat{\mathfrak{G}}^{-1}(\textbf{r},\textbf{r}';t-t')=\left(
  \begin{array}{cc}
    i\partial_t-\frac{\hat{\textbf{p}}^2}{2m}+\mu-u(\textbf{r})-(U_0-2U_1)|\psi_0(\textbf{r})|^2 & 0 \\
    0 & -i\partial_t-\frac{\hat{\textbf{p}}^2}{2m}+\mu-u(\textbf{r})-(U_0-2U_1)|\psi_0(\textbf{r})|^2 \\
  \end{array}
\right)\delta_{\textbf{r},\textbf{r}'}\delta_{t,t'},\\
\label{eq11}
\hat{G}^{-1}(\textbf{r},\textbf{r}';t-t')
=\left(
  \begin{array}{cc}
    i\partial_t-\frac{\hat{\textbf{p}}^2}{2m}+\mu-u(\textbf{r})-2U_0|\psi_0(\textbf{r})|^2 & -U_0|\psi_0(\textbf{r})|^2 \\
    -U_0|\psi_0(\textbf{r})|^2 & -i\partial_t-\frac{\hat{\textbf{p}}^2}{2m}+\mu-u(\textbf{r})-2U_0|\psi_0(\textbf{r})|^2 \\
  \end{array}
\right)\delta_{\textbf{r},\textbf{r}'}\delta_{t,t'}.
\end{gather}
\end{widetext}
From Eq.~(\ref{eq9}) it is evident that the TE-TM mixing results in transitions between the circularly-polarized components of the EP spinor, as expected. The formal solution of the system of Eqns.~(\ref{eq9}) reads:
\begin{gather}\label{eq12}
	\left(
         \begin{array}{c}
           \delta\psi_+ (\textbf{r},t)\\
           \delta\psi_+^*(\textbf{r},t) \\
         \end{array} \right)
         =\iint d\textbf{r}'dt'\hat{G}^R(\textbf{r},\textbf{r}';t-t')\times~~~~~~~~~~~\\
         \nonumber
\times\left[\psi_0(\textbf{r}')\mathcal{F}(\textbf{r}',t')
\left(
                                          \begin{array}{c}
                                            1 \\
                                            1 \\
                                          \end{array}
                                        \right)+\hat{K}	
                                        \left(
         \begin{array}{c}
           \delta\psi_-\\
           \delta\psi_-^*\\
         \end{array} \right)\right],\\\nonumber
	\left(
         \begin{array}{c}
           \delta\psi_- (\textbf{r},t)\\
           \delta\psi_-^*(\textbf{r},t) \\
         \end{array} \right)
=\iint d\textbf{r}'dt'\hat{\mathfrak{G}}^R(\textbf{r},\textbf{r}';t-t')\hat{K}^*
	\left(
         \begin{array}{c}
           \delta\psi_+\\
           \delta\psi_+^*\\
         \end{array} \right),
\end{gather}
and now the components of the spin density can be expressed via the first-order corrections to the EP spinor:
\begin{gather}\label{eq13}
S^{x}(\textbf{r},t)\approx \langle\psi_0(\textbf{r})[\delta\psi_-(\textbf{r},t)+\delta\psi_-^*(\textbf{r},t)]\rangle,\\\nonumber
S^{y}(\textbf{r},t)\approx -i\langle\psi_0(\textbf{r})[\delta\psi_-(\textbf{r},t)-\delta\psi_-^*(\textbf{r},t)]\rangle,\\\nonumber
S^z(\textbf{r},t)-\langle\psi_0^2(\textbf{r})\rangle\approx\langle\psi_0(\textbf{r})[\delta\psi_+(\textbf{r},t)+\delta\psi_+^*(\textbf{r},t)]\rangle.
\end{gather}
Let us consider different regimes.
%
%
%
%
%
%

\textit{Ballistic transport.---}
In an ideal case of pure sample where polariton-impurity scattering processes can be neglected, the ground state (condensate) wave function, $\psi_0(\textbf{r})$, is uniform in space.
%
%
%
\begin{figure}[!t]
	\includegraphics[height=0.38\linewidth]{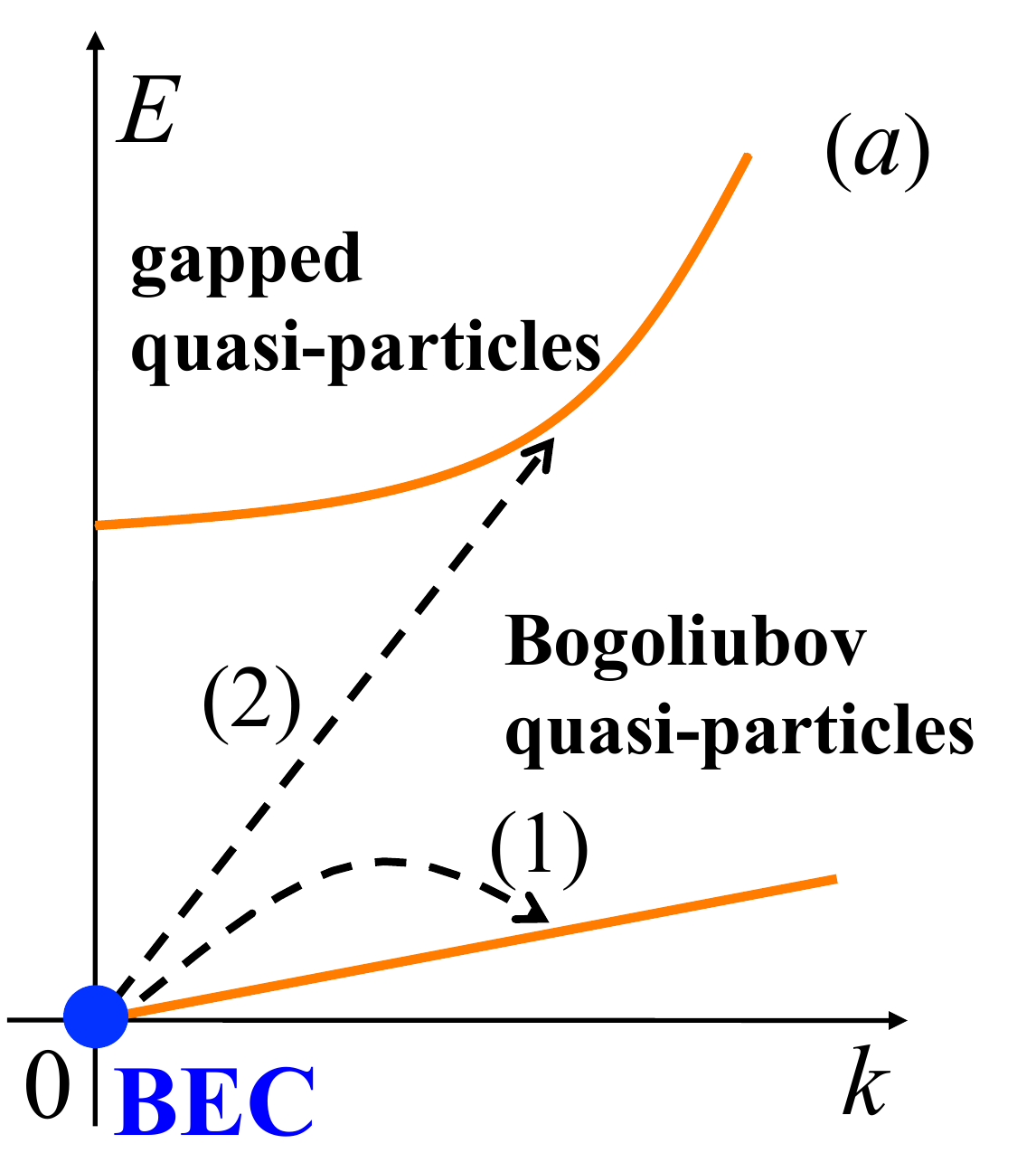}
	\includegraphics[height=0.40\linewidth]{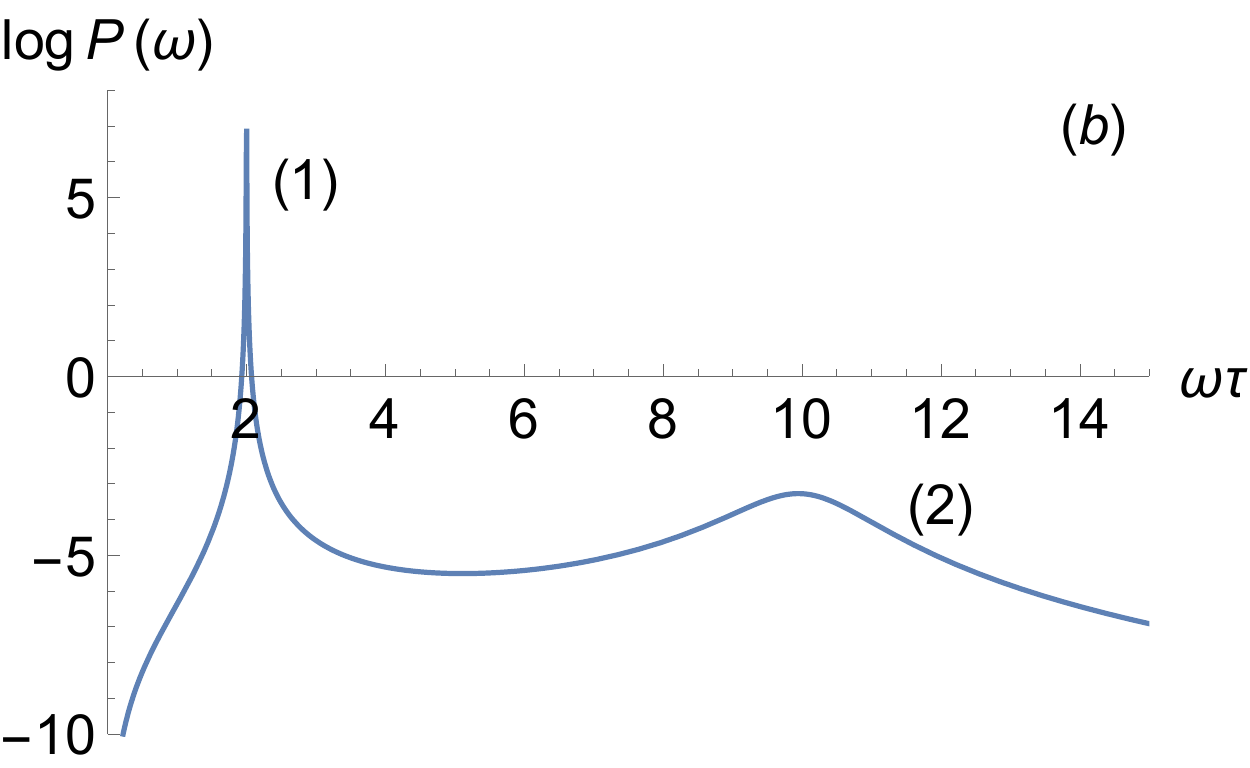}
	\caption{(a) Schematic of the quasi-particle spectrum of the system with two types of transitions: (1) and (2). Blue solid dot is the condensate of `+' polarized EPs. (b) Power absorption spectrum in semi-log scale. The peaks (1) and (2) result from the transitions (1) and (2) from (a).}
	\label{Fig1}
\end{figure}
%
%
%
Then from Eq.~(\ref{eq8}) we get $\psi_0(\textbf{r})\equiv\psi_0=\sqrt{n_c}$ and $\mu=U_0n_c$, and we can calculate the Green's functions:
\begin{gather}\label{eq14}
\hat{\mathfrak{G}}^R(\epsilon,\textbf{p})=\frac{\left(
  \begin{array}{cc}
    \epsilon+\mathcal{E}_p & 0 \\
    0 & -\epsilon+\mathcal{E}_p \\
  \end{array}
\right)}{(\epsilon+i\delta)^2-\mathcal{E}_p^2},
\\\nonumber
\hat{G}^R(\epsilon,\textbf{p})
=\frac{\left(
  \begin{array}{cc}
    \epsilon+E_p+U_0n_c & -U_0n_c \\
    -U_0n_c & -\epsilon+E_p+U_0n_c \\
  \end{array}
\right)}{(\epsilon+i\delta)^2-\epsilon_p^2},
\end{gather}
where $\epsilon_p=\sqrt{E_p(E_p+2U_0n_c)}=sp\sqrt{1+p^2\xi^2}$ is a Bogoliubov quasiparticle spectrum, $\xi=1/2Ms$ is a healing length, $s^2=U_0n_c/M$ is a Bogoliubov excitations velocity and $\mathcal{E}_p=2|U_1|n_c+E_p$ is a gapped dispersion branch of low-populated EP circular component~\cite{RuboBookChapter}, see Fig.~\ref{Fig1}a.
Then the exact solutions of Eq.~(\ref{eq9}) are
\begin{gather}\label{eq15}
\left(
  \begin{array}{c}
    \delta\psi_+(\textbf{k},\omega) \\
    \delta\psi^*_+(\textbf{k},\omega) \\
  \end{array}
\right)
=\sqrt{n_c}\hat{L}^{-1}(\textbf{k},\omega)\mathcal{F}(\textbf{k},\omega)
\left(
  \begin{array}{c}
    1 \\
    1 \\
  \end{array}
\right),\\\nonumber
\left(
  \begin{array}{c}
    \delta\psi_-(\textbf{k},\omega) \\
    \delta\psi^*_-(\textbf{k},\omega) \\
  \end{array}
\right)=\hat{\mathfrak{G}}^R(\textbf{k},\omega)\hat{K}^*\left(
  \begin{array}{c}
    \delta\psi_+(\textbf{k},\omega) \\
    \delta\psi^*_+(\textbf{k},\omega) \\
  \end{array}
\right)
\end{gather}
with $\hat{L}^{-1}=(\hat{G}^{-1}-\alpha^2p^4\mathfrak{G})^{-1}$. Calculating this inverse matrix, we keep all the $\alpha$-containing terms in the numerator and disregard their contribution to the denominator (in determinant which appears in the matrix calculation), assuming that the TE-TM splitting is small and does not affect the dispersions, $\epsilon_k$ and $\mathcal{E}_k$.
Then in the lowest order in $\alpha$ we obtain the transverse,
\begin{gather}\label{eq16}
\chi_{xz}(\textbf{k},\omega)=\frac{\alpha n_cg_s}{2\mu_b^{-1}}
\frac{A_++A_-}{D_{\mathcal{E}}D_\epsilon},\\
\label{eq16_2}
\chi_{yz}(\textbf{k},\omega)=\frac{\alpha n_cg_s}{i2\mu_b^{-1}}\frac{A_+-A_-}{{D_{\mathcal{E}}D_\epsilon}},
\end{gather}
where $A_+=k_+^2(\omega+E_k)(\omega+\mathcal{E}_k)$, $A_-=k_-^2(\omega-E_k)(\omega-\mathcal{E}_k)$, $D_{\mathcal{E}}=(\omega+i\delta)^2-\mathcal{E}_k^2$, $D_\epsilon=(\omega+i\delta)^2-\epsilon_k^2$, and longitudinal,
\begin{gather}\label{eq17}
\chi_{zz}(\textbf{k},\omega)=\frac{g_s\mu_bn_cE_k}{D_\epsilon}
\left[1+\frac{(2M\alpha)^2E_k\mathcal{E}_k}{D_{\mathcal{E}}}\right],
\end{gather}
pseudo-spin susceptibilities.
From these expressions it is evident that the TE-TM coupling results in a non-zero transverse spin polarization response of the EP gas and a correction to the longitudinal susceptibility. It experiences resonance in the vicinity of the frequency of the collective (Bogoliubov) mode of the condensate, $\omega\approx \epsilon_k$.
Moreover, TE-TM splitting results in transitions of particles between the spin-polarized components of the EP doublet which results in emergence of an additional resonance at $\omega\approx \mathcal{E}_k$.
Similarly, the transverse susceptibilities also demonstrate a double-resonance structure determined by the longitudinal-transverse correction to EP spectrum. It should be mentioned that both the transverse
\eqref{eq16},~\eqref{eq16_2} and longitudinal \eqref{eq17} susceptibilities diverge at frequencies corresponding to the exact resonance, $\omega=\epsilon_k$ or $\omega=\mathcal{E}_k$. It happens due to the infinitely small scattering rates of `$+$' and `$-$' EPs.

%
%
%
%
%
%
\textit{Finite polariton-impurities scattering.---}
Accounting for the scattering mechanisms results in the line broadening and finite values of susceptibilities \eqref{eq16}-\eqref{eq17} at resonances. The most significant contributions to EP non-radiative lifetime at low temperatures are given by the polariton-polariton~\cite{RefKovalevSavenkoIorsh} and polariton-disorder scattering. We will analyze here the latter case considering the disorder caused by impurities.
A naive approach, commonly used in literature, is to assume that the $i\delta$ terms in \eqref{eq16}, \eqref{eq16_2} and \eqref{eq17} have finite value, associated with some phenomenological particle scattering time, $\delta\rightarrow 1/\tau$, where $\tau$ is independent of the momentum and energy. However, an important question is, will it remain so when EPs are in the condensed state?

In the presence of a disorder, the ground state of the system is to be determined from Eq.~(\ref{eq8}). To solve this equation and find $\psi_0(\textbf{r})$, we follow the approach suggested in~\cite{RefSuris} for a 3D excitonic system. In its framework, the impurity field, $u(\textbf{r})$, produces a static fluctuation of the condensate density, $\psi_0(\textbf{r})$. However, it is assumed to be weak enough thus it cannot destroy the condensate.
Then we can write: $\psi_0(\textbf{r})=\sqrt{n_c}+\phi(\textbf{r})$, where $|\phi(\textbf{r})|\ll\sqrt{n_c}$ is a correction.
Further, linearization of Eq.~(\ref{eq8}) with respect to $\phi(\textbf{r})$ gives:
%
\begin{gather}\label{eq18}
[\hat{E}_{\textbf{p}}-\delta\mu+2U_0n_c+u(\textbf{r})]\frac{\phi(\textbf{r})}{\sqrt{n_c}}=-\left(u(\textbf{r})-\delta\mu\right),
\end{gather}
where $\delta\mu=\mu-U_0n_c$ is a correction to the chemical potential. The formal solution of this equation reads:
\begin{gather}\label{eq19}
\phi(\textbf{r})=\sqrt{n_c}\int d\textbf{r}'g(\textbf{r},\textbf{r}')\left(u(\textbf{r}')-\delta\mu\right),
\end{gather}
where
\begin{gather}\label{eqGreen}
[-\hat{E}_{\textbf{p}}+\delta\mu-2U_0n_c-u(\textbf{r})]g(\textbf{r},\textbf{r}')=\delta(\textbf{r}-\textbf{r}')
\end{gather}
and $\delta\mu$ is determined by the condition $\langle\phi(\textbf{r})\rangle=0$. In the lowest order of the perturbation theory, we use the Green's function, $g(\textbf{r},\textbf{r}'),$ taken at $u(\textbf{r})=0$ and find the fluctuating part of the ground state wave function:
\begin{gather}\label{eqFluctGround}
\phi(\textbf{p})=\sqrt{n_c}g(\textbf{p})u(\textbf{p}),\,\,\,g(\textbf{p})=-\frac{1}{2U_0n_c}\frac{1}{1+p^2\xi^2}
\end{gather}
and $\delta\mu=0$. Now one can find the disorder-averaged Green's functions and EP-impurity scattering times.
To do this, one needs to linearize Eq.~(\ref{eq11}) with respect to $\phi(\textbf{r})$ to get the matrix
equations: $\hat{G}^R=\hat{G}^R_0+\hat{G}^R_0\hat{X}\hat{G}^R$ and
$\hat{\mathfrak{G}}^R=\hat{\mathfrak{G}}^R_0+\hat{\mathfrak{G}}^R_0\hat{\mathcal{X}}\hat{\mathfrak{G}}^R$,
where the bare (without disorder) functions, $\hat{G}^R_0$, $\hat{\mathfrak{G}}^R_0$, are given by Eq.~(\ref{eq14}) and we denote
\begin{gather}
\label{eqPotentials}
\hat{X}(\textbf{r})=u(\textbf{r})\left(
                                   \begin{array}{cc}
                                     1 & 0 \\
                                     0 & 1 \\
                                   \end{array}
                                 \right)+
2U_0\sqrt{n_c}\phi(\textbf{r})
			\left(
                      \begin{array}{cc}
                        2 & 1 \\
                        1 & 2 \\
                      \end{array}
                    \right),\\
\label{eqPotentials2}
\hat{\mathcal{X}}(\textbf{r})=[u(\textbf{r})+2\sqrt{n_c}(U_0-2U_1)\phi(\textbf{r})]
\left(
                      \begin{array}{cc}
                        1 & 0 \\
                        0 & 1 \\
                      \end{array}
                    \right).
\end{gather}
These potentials describe the EP scattering on impurity field (terms $\sim u(\textbf{r})$) and on the static fluctuations of the condensate density (terms $\sim\phi(\textbf{r})$). Now we apply a standard Feynman diagram technique and show that in the lowest order of the Born approximation, the impurity self-energies can be written in the standard form: $\hat{W}(\textbf{r}-\textbf{r}')=\langle \hat{X}(\textbf{r})\hat{G}^R_0(\textbf{r}-\textbf{r}')\hat{X}(\textbf{r}')\rangle$ and
$\hat{\mathcal{W}}(\textbf{r}-\textbf{r}')=\langle \hat{\mathcal{X}}(\textbf{r})\hat{\mathfrak{G}}^R_0(\textbf{r}-\textbf{r}')\hat{\mathcal{X}}(\textbf{r}')\rangle$.
The Green's functions averaged over the disorder can be found from the matrix Dyson equations \cite{chkov},
$\langle\hat{G}^{-1}\rangle=\hat{G}^{-1}_0-\hat{W}$ and $\langle\hat{\mathfrak{G}}^{-1}\rangle=\hat{\mathfrak{G}}^{-1}_0-\hat{\mathcal{W}}$.
At this point, the general consideration with the spectrum of the Bogliubov quasiparticles,
$\epsilon_k=sk\sqrt{1+k^2\xi^2}$ and arbitrary $k$, becomes a tricky issue. However, we do not have to find the general solution since we can restrict our consideration to
the most important case of quasi-linear Bogliubov dispersion, $\epsilon_k\approx sk$, which is hold
under the condition $k\xi\ll1$. This case can be treated analytically. Taking into account Eqs.~\eqref{eqPotentials} and~\eqref{eqPotentials2}, we find:
\begin{gather}\label{eq22}
\hat{\mathcal{W}}(\epsilon)=u_0^2\left(\frac{2U_1}{U_0}\right)^2\int \frac{d\textbf{p}}{(2\pi)^2}\hat{\mathfrak{G}}^R_0(\textbf{p},\epsilon),\\\nonumber
\hat{W}(\epsilon)=u_0^2\int \frac{d\textbf{p}}{(2\pi)^2}\left(
                                                          \begin{array}{cc}
                                                            1 & 1 \\
                                                            1 & 1 \\
                                                          \end{array}
                                                        \right)
\hat{G}^R_0(\textbf{p},\epsilon)\left(
                                                          \begin{array}{cc}
                                                            1 & 1 \\
                                                            1 & 1 \\
                                                          \end{array}
                                                        \right).
\end{gather}
Substituting the bare Green's functions~(\ref{eq14}) into Eq.~(\ref{eq22}),
averaging over the disorder and using the matrix equations $\langle\hat{G}^{-1}\rangle=\hat{G}^{-1}_0-\hat{W}$,
$\langle\hat{\mathfrak{G}}^{-1}\rangle=\hat{\mathfrak{G}}^{-1}_0-\hat{\mathcal{W}}$,
we can now find the impurity-mediated scattering times.


\textit{Results and discussion.---}
In our chosen limit $k\xi\ll1$ (in practice, $k\xi<1$) and at the mass shell $\epsilon=sk$ for `$+$'
polarized polaritons and $\epsilon=\mathcal{E}_k$ for `$-$' polaritons, we find the polariton-impurity scattering rates:
\begin{gather}\label{eq23}
\gamma_k^+=\frac{1}{\tau}(k\xi)^3,\,\,\,\,
\gamma_k^-=\frac{1}{\tau}\left(\frac{2U_1}{U_0}\right)^2.
\end{gather}
Here $1/\tau=Mu_0^2$ is the inverse scattering time in the normal (not condensed) state. As it is expected to be, `$-$' polaritons which are assumed to be in the normal state, have regular scattering lifetime ($2U_1/U_0\sim 1$), whereas the scattering of the polaritons in the condensed state turns out severely suppressed due to $(k\xi)^3\ll1$.

Scattering rates~\eqref{eq23} together with the expressions for the longitudinal and transverse spin susceptibilities,~\eqref{eq16}-\eqref{eq17}, are the key results of this Letter. They determine the paramagnetic absorption line widths. From these expressions it is obvious that the response line width of the macroscopically occupied component of the polariton function (in our case it is `$+$' component) is much less in comparison with the line width of the initially unoccupied `$-$' component of the doublet, since $\gamma_k^+/\gamma_k^-\sim (k\xi)^3\ll1$. This fundamental result can also be used in analysing experiments, whether one of the components is in the Bose-condensed state or not.


The response of the system is conventionally described by the  power absorption which is proportional to the imaginary part of longitudinal susceptibility,
\begin{equation}
P_{k\omega}\sim - \omega\, \textmd{Im}\,\chi_{zz}(\textbf{k},\omega).
\end{equation}
To explain qualitatively the structure of its spectrum, we consider the quantum transitions of the particles under external perturbation, shown in Fig.~\ref{Fig1}a. In usual electronic systems, the power absorption spectrum of the paramagnetic resonance is characterised by single resonance associated with the transitions between two spin-resolved electron levels. In contrast to this situation, in our bosonic system we have a double-peak structure of the resonance. This is due to the fact that effectively our system has three levels. Indeed, as one can see from Fig.~\ref{Fig1}a, beside the condensate itself there are two branches of excitations with energies $\epsilon_k$ and ${\cal E}_k$ in the system. The transitions from the BEC to these two branches results in the double resonance structure, see Fig.~\ref{Fig1}b. Thus the presence of the BEC is crucial for the considered effect.

The second important difference between the considered effect and the regular paramagnetic resonance is the requirement to use nonuniform alternating magnetic field instead of a homogeneous one. In other words, finite values of $k=|\textbf{k}|$ are required. The reason is that EPs in BEC have zero momentum and in order to excite them one has to transfer the momentum from an external excitation.

\begin{figure}[!t]
	\includegraphics[height=0.3\linewidth]{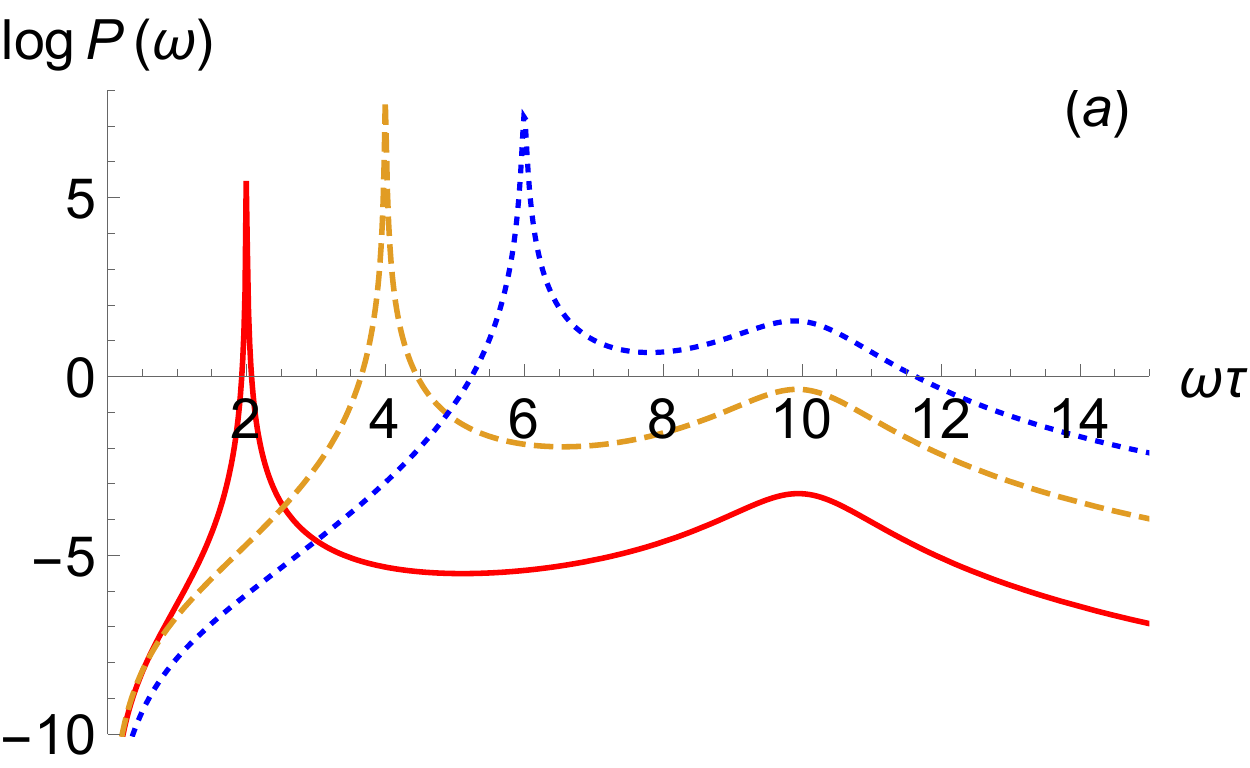}
	\includegraphics[height=0.3\linewidth]{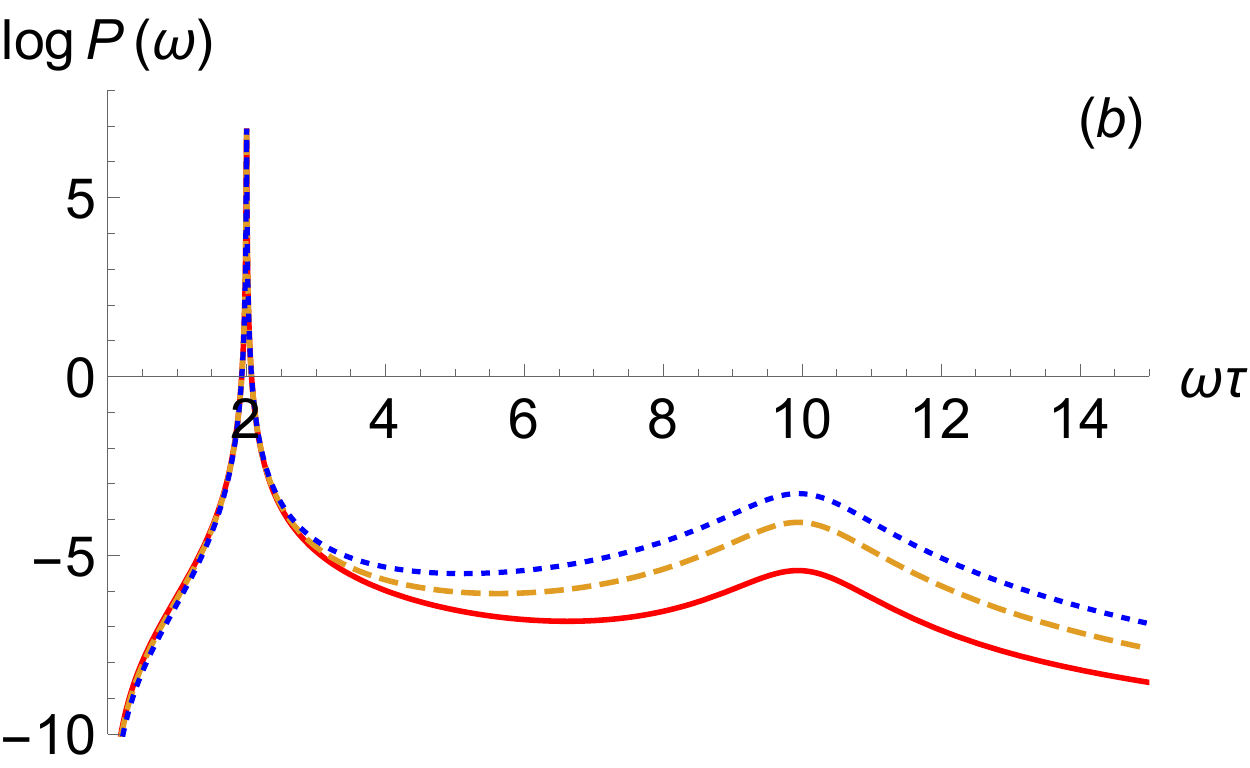}
	\caption{Power absorption spectrum in semi-log scale for (a) various values of $k\xi$: $0.1$ (red solid), $0.2$ (yellow dashed) and $0.3$ (blue dotted) and (b) various values of $M\alpha$: $0.1$ (red solid), $0.2$ (yellow dashed) and $0.3$ (blue dotted curve).}
	\label{Fig2}
\end{figure}

In our theory we operate with two free parameters which can be determined by the experiment and the semiconductor sample: (i) the wave vector of the external perturbation, $k$, and (ii) impurity scattering time, $\tau$. For (i), we have the following constraint: $k\xi\ll1$.
In order to fix (ii), we take $U_0n\tau=10$, since our theory is feasible if $U_0n\tau\gg1$. Taking into account that $|U_1|\approx 0.5U_0$, we have $2|U_1|n\tau=10$. Using the dimensionless units of TE-TM splitting, $M\alpha$, we plot the power absorption spectrum in Fig.~\ref{Fig2} for different values of $k\xi$ (a) and  $M\alpha$ (b). Clearly, both the positions of the resonances and their widths depend on (i) and (ii). It can be useful for experimental testing of our theory. The value $k$ determines the position and width of the first resonance, whereas the value of TE-TM splitting, $\alpha$, determines the height of the second resonance. In fact, the position of the second resonance is determined by the EP blueshift value, $2|U_1|n_c$. This value also gives an estimation of the characteristic magnetic field frequency, $\omega\sim 2|U_1|n_c\sim 10^{11} s^{-1}$, required to observe the effect. The first resonance occurs at the frequency $\omega\sim sk$ and is thus determined by the value of $k$. 


\textit{Outlook.---}
One more important point to mention is the role of polariton-polariton scattering to the widths of peaks of the paramagnetic resonance. It can become significant in a particularly clean cavity, where impurity scattering is negligible.
It is known that the particle-particle scattering rate in a 2D Bose gas calculated within the Bogliubov theory depends on the wave vector as $k^3$. One can expect that the particle-particle scattering rate in the normal (not Bose-condensed) phase will behave as a square of its energy, $E_k^2\sim k^4$ and it will be less than in the condensed phase. Thus we expect that in this situation, the width of the low-occupied component can become narrower than the macroscopically occupied component which is the opposite situation to what we have observed here. In order to give a conclusive answer, one should also consider the scattering between the condensed, $\psi_+$, and non-condensed, $\psi_-$, EPs. This interesting question is beyond the scope of  present Letter.

The second issue is the case $U_1>0$. In the case of equally populated circular components of the EP doublet, occurring at $U_1>0$, the Zeeman splitting becomes strongly suppressed by the particle-particle interaction up to some critical value of the constant magnetic field~\cite{RefShelykhSpin,RefNature}. Thus, the paramagnetic resonance may only occur if the magnitude of the alternative magnetic field exceeds some critical value. This question also deserves an extra consideration.

Finally, we believe that a similar physics might be observed in indirect exciton gases with spin-orbit Rashba or Dresselhaus interaction in the limit of large exchange interaction between the electron and hole within the exciton. Indeed, as it has been shown in~\cite{Durnev}, the indirect exciton Hamiltonian has a form which exactly coincides with the EP Hamiltonian in the presence of the TE-TM splitting.


\textit{Conclusions.---}
We have developed a microscopic theory of paramagnetic resonance in a spin-polarized polariton gas in a disordered microcavity. Pseudospin susceptibilities were calculated accounting for TE-TM splitting. We have shown that both longitudinal and transverse susceptibilities have a double resonance structure, responsible for different polariton spin states, and calculated the widths of the peaks of the paramagnetic resonance  taking into account the polariton-impurity scattering. In contrast to ordinary disordered electronic systems, exciton polaritons in the presence of the BEC phase can scatter off both the impurity potential and impurity-stimulated fluctuations of the condensate density. We analyze those scattering processes and find that the polariton-impurity scattering rates are dramatically different for macroscopically, on one hand, and low occupied, on the other hand, components of the polariton doublet.


\emph{Acknowledgments.---} We thank A. Chaplik for discussions and critical reading of the manuscript.
V.M.K acknowledges the support from RFBR grant $\#16-02-00565a$.
I.G.S. acknowledges support of the Project Code (IBS-R024-D1), Australian Research Council Discovery Projects funding scheme (Project No. DE160100167), President of Russian Federation (Project No. MK-5903.2016.2), and Dynasty Foundation. V.M.K. also thanks the IBS Center of Theoretical Physics of Complex Systems in Korea for hospitality.


\end{document}